\documentclass{article}

\usepackage{spconf}
\usepackage{graphicx} % Required for inserting images
\usepackage{amsmath, amssymb, amsfonts}
\usepackage{booktabs}
\usepackage{multirow}
\usepackage{pifont}
\usepackage{url}

\def\W{{\mathbf W}}
\def\x{{\mathbf x}}
\def\ii{{\hat{\imath}}}	% Definition of the imaginary symbol i
\def\bH{{\mathbb H}}
\def\bC{{\mathbb C}}
\def\bQ{{\mathbb Q}}

\title{Hypercomplex Multimodal Emotion Recognition from EEG and Peripheral Physiological Signals}
\name{Eleonora Lopez, Eleonora Chiarantano, Eleonora Grassucci, and Danilo Comminiello}
\address{Dept. of Information Eng., Electronics and Telecom., Sapienza University of Rome, Italy}

\begin{document}
% \ninept
\maketitle

\begin{abstract}

% General introductory sentence about the main subject emerging from the
% title (model/application/problem).
% 2. Typical general problems related to the subject introduced in the previous
% sentence.
% 3. Specific problems that make the subject difficult to deal with, challenging,
% and/or still an open problem.
% 4. Describe how we want to address and solve the above problems.
% Highlight any novelty.
% 5. Sentence on experimental results with respect to the SOTA.
% 6. Add a github link with our code repository.

Multimodal emotion recognition from physiological signals is receiving an increasing amount of attention due to the impossibility to control them at will unlike behavioral reactions, thus providing more reliable information. Existing deep learning-based methods still rely on extracted handcrafted features, not taking full advantage of the learning ability of neural networks, and often adopt a single-modality approach, while human emotions are inherently expressed in a multimodal way. In this paper, we propose a hypercomplex multimodal network equipped with a novel fusion module comprising parameterized hypercomplex multiplications. Indeed, by operating in a hypercomplex domain the operations follow algebraic rules which allow to model latent relations among learned feature dimensions for a more effective fusion step. We perform classification of valence and arousal from electroencephalogram (EEG) and peripheral physiological signals, employing the publicly available database MAHNOB-HCI surpassing a multimodal state-of-the-art network. The code of our work is freely available at \url{https://github.com/ispamm/MHyEEG}.

\end{abstract}

\section{Introduction}
\label{sec:intro}

Emotion is an essential part of human communication that plays a vital role in the overall quality and outcome of interactions. Thus, automatic emotion recognition and affective computing have gained much interest, also considering the wide range of applications in human-computer interaction (HCI) \cite{wu2022novel}. Humans manifest emotions in a multimodal way, including facial expressions, speech, body language, and physiological signals. While behavioral reactions can be easily controlled, for example, real emotion can be concealed by adjusting expressions or tone of voice, physiological signals cannot be governed at will, thus being more reliable for recognizing human emotion \cite{zhang2022multimodal}. Therefore, on account of the development of non-invasive and inexpensive wearable devices, physiological-based emotion recognition has become a hot topic in affective computing research. Among these, electroencephalography (EEG) is a measure of the electrical activity of the brain that is directly correlated with the cognitive process and can provide key information regarding emotional states being characterized by excellent temporal resolution \cite{zhang2022ganser}. Therefore, EEG-based analysis has received an increasing amount of attention for a variety of applications such as epileptic seizure detection \cite{boonyakitanont2020seizure}, general EEG classification \cite{tan2018dtransfer} and emotion recognition \cite{li2021hierarchical}.  

Nevertheless, most studies do not take full advantage of the learning ability of deep learning models and most of the time focus on a single-modality approach. In fact, the input to the neural model is often extracted features instead of the raw data and corresponding to a single modality, generally EEG, when in reality human emotions are intrinsically multimodal, with different modalities describing different aspects of an emotional reaction and correlations among them providing critical information if exploited correctly \cite{zhang2022multimodal}. Recent works have started to take a multimodal approach, but most rely on trivial techniques and few studies explore more emerging paradigms such as multimodal learning \cite{stahlschmidt2022multimodal}. Therefore, effectively learning from multiple physiological signals to produce more powerful feature representations is still an open problem. Motivated by the described challenges, in this paper we address the more difficult approach of learning directly from raw signals and propose a multimodal architecture with a novel fusion module that exploits the properties of algebras in the hypercomplex domain to truly take advantage of correlations characteristic of EEG and peripheral physiological signals. 

Parameterized hypercomplex neural networks (PHNNs) are an emerging family of models which operate in a hypercomplex number domain \cite{zhang2021phm, grassucci2021phnns}. They have been introduced in order to generalize the more common quaternion neural networks (QNNs) which are defined in the quaternion domain and are thus limited to $4$D input data but possess very powerful capabilities \cite{parcollet2019survey}. In fact, thanks to quaternion algebra operations, such as the Hamilton product, these models are endowed with the ability to capture not only global relations as any neural network but also local relations among input data, unlike real-valued counterparts, as well as being more lightweight \cite{parcollet2019qcnn}. Thanks to the introduction of parameterized hypercomplex multiplication (PHM) and convolution (PHC), these advantages have been extended to inputs of any dimensionality $n$, with a reduction of parameters of $1/n$.

Owing to these advantages, we design a hypercomplex multimodal network with a novel fusion module defined in the hypercomplex domain, thus comprising PHM layers that thanks to hypercomplex algebra properties endow the architecture with the ability to model correlations among the learned latent features, thus learning a more effective fused representation. Specifically, we perform classification of valence and arousal from EEG, electrocardiogram (ECG), galvanic skin response (GSR), and eye data, and we validate the proposed approach on a publicly available benchmark, MAHNOB-HCI \cite{soleymani2011mahnob}, showing how our method outperforms a multimodal state-of-the-art network.

\section{Background}
\label{sec:background}

A plethora of machine learning approaches for emotion recognition have been proposed \cite{he2022crossday, liu2017realtime, martinez2021exploring}. However, employing such methods requires extensive domain knowledge to extract relevant features. On the other hand, deep learning models are able to learn features directly from the raw data and thus learn a powerful latent representation. Due to these advantages, many deep learning-based methods have been investigated \cite{maeng2020deep, du2022efficient, wang2019design, rayatdoost2018cross}. Nonetheless, even though such works employ neural networks, they still rely on extracted features, such as power spectral density (PSD) and differential entropy (DE), instead of taking full advantage of the representational learning ability of neural models. Rather, a study that employs raw EEG signals has proposed a $3$D representation of the data to be processed by a $3$D convolutional neural network (CNN) \cite{salama2018eeg}. However, all aforementioned methods focus on a single-modality approach which is suboptimal \cite{zhang2022multimodal}. Thus in order to exploit the information contained in different modalities, recent studies adopt a multimodal approach for emotion recognition, some still relying on extracted features \cite{zhang2022multimodal, tan2020fusionsense, rayatdoost2020expression} and very few that directly employ raw data \cite{zeng2020elderly, nakisa2020automatic, dolmans2021workload}, where the latter focuses on perceived mental workload classification instead of emotion recognition.

Aside from feature extraction, the crucial step of multimodal learning is the fusion strategy. Surely, much of the research in this field has focused on this aspect, as there are a multitude of manners to incorporate information from different modalities. Starting from the most trivial, i.e. early fusion in which data from different modalities is concatenated to form a single input to the model, and late fusion which consists in aggregating decisions of different networks trained separately on each modality to obtain a final output. Both strategies suffer from several problems, where the first does not take into account the different nature of the input modalities, not taking advantage of complementary information and not allowing to identify relations among them, while the second does not exploit cross-modal information during learning at all, also requiring to optimize a different network for each modality. Instead, more complex strategies fall under the name of intermediate fusion, which consists in first learning modality-specific latent representations that are subsequently fused together for further processing \cite{stahlschmidt2022multimodal}. Thus, in this paper, we investigate and propose a novel technique that allows to effectively grasp correlations between the different modalities during learning thanks to its definition in a hypercomplex algebraic system.

\section{Methodology}
\label{sec:method}

% \begin{itemize}
%     \item This section must represent the core of your paper.
%     \item Theoretical novelty must be clearly highlighted.
%     \item If possible add a scheme of your method to help the comprehension.
%     \item If possible add some additional insights that derive from the proposed method (e.g., theoretical analyses, ad hoc regularization, feature selection, specific layers,...)
%     \item Length: 1 page and a half
% \end{itemize}

\begin{figure*}[t]
    \centering
    \includegraphics[width=\textwidth]{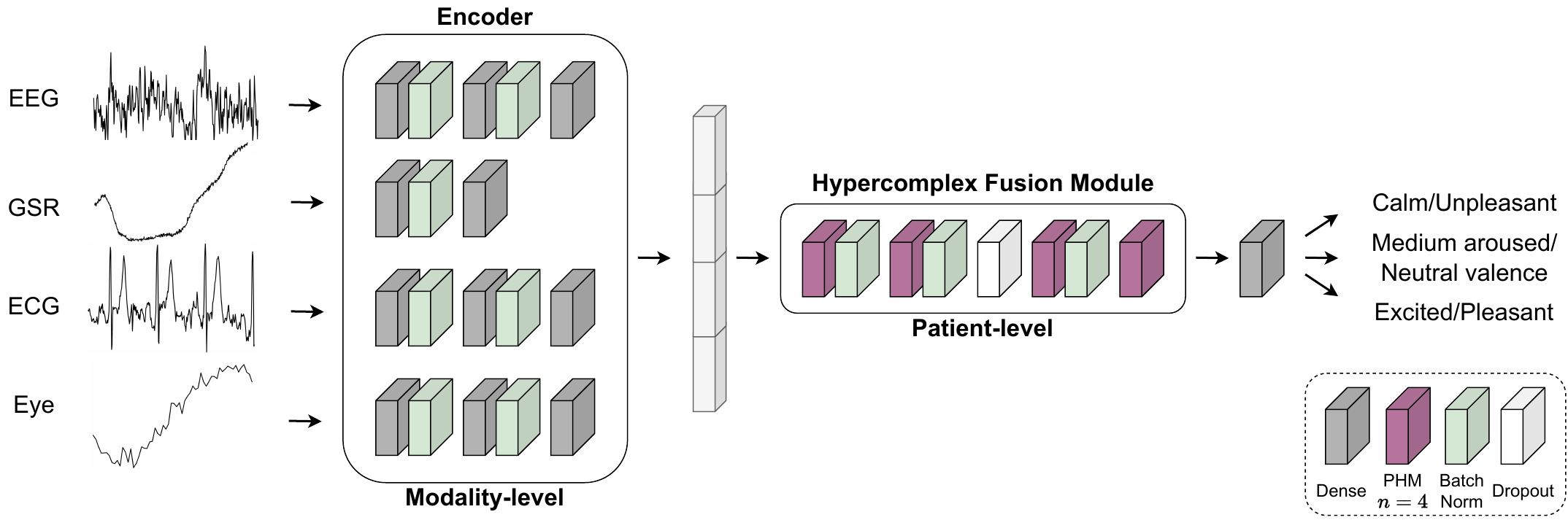}
    \caption{HyperFuseNet architecture. The encoder learns modality-specific latent representations in the real domain, which are then merged together and processed by the novel fusion module in the hypercomplex domain with $n=4$. Finally, a fully-connected layer produces the prediction for arousal/valence.}
    \label{fig:method}
\end{figure*}

\subsection{Hypercomplex neural models}

Hypercomplex neural networks are neural models defined in a hypercomplex number system $\bH$ which is regulated by the respective algebra rules that define addition and multiplication operations. A generic hypercomplex number is defined as 

\begin{equation}
    h = h_0 + h_i \ii_i + \ldots + h_n \ii_n, \qquad i=1, \ldots, n
\label{eq:hyp_num}
\end{equation}

\noindent where $h_0, \ldots, h_n \in \mathbb R$ and $\ii_i, \ldots, \ii_n \in \bH$ are the imaginary units.

The general hypercomplex domain $\bH$ includes various algebraic systems such as the complex $\bC$ domain when $n=2$ and the quaternion $\bQ$ domain when $n=4$, where quaternion neural networks (QNNs) operate in. In fact, algebra rules are defined only at predefined dimensions of $n=2^m$, with $m \in \mathbb N$, owing to the fact that hypercomplex algebras are included in the family of Cayley-Dickson algebras. Thus, each of these number systems is identified by the number of imaginary units and consequently by the different definitions of the multiplication operation as a result of the disparate interactions among imaginary units. For example, in the quaternion domain, the product is non-commutative, with $\ii_1 \ii_2 \neq \ii_2 \ii_1$. Therefore, in the latter domain, the Hamilton product was introduced, which also regulates the matrix multiplication in fully connected layers and the convolution operation in convolutional layers, since both the weight matrix and the input are encapsulated into a quaternion in the following way: $\W = \W_0 + \W_1 \ii_1 + \W_2 \ii_2 + \W_3 \ii_3$ and $\x = \x_0 + \x_1\ii_1 + \x_2\ii_2 + \x_3\ii_3$, respectively. As a consequence, the matrix multiplication of a general fully connected layer becomes

\begin{equation}
{\bf{W}}{\bf{x}} = \left[ {\begin{array}{*{20}c}
   \hfill {{\bf{W}}_0 } & \hfill { - {\bf{W}}_1 } & \hfill { - {\bf{W}}_2 } & \hfill { - {\bf{W}}_3 } \\
   \hfill {{\bf{W}}_1 } & \hfill {{\bf{W}}_0 } & \hfill { - {\bf{W}}_3 } & \hfill {{\bf{W}}_2 } \\
   \hfill {{\bf{W}}_2 } & \hfill {{\bf{W}}_3 } & \hfill {{\bf{W}}_0 } & \hfill { - {\bf{W}}_1 } \\
   \hfill {{\bf{W}}_3 } & \hfill { - {\bf{W}}_2 } & \hfill {{\bf{W}}_1 } & \hfill {{\bf{W}}_0 } \\
\end{array}} \right] \left[ {\begin{array}{*{20}c}
   {{\bf{x}}_0 } \hfill  \\
   {{\bf{x}}_1 } \hfill  \\
   {{\bf{x}}_2 } \hfill  \\
   {{\bf{x}}_3 } \hfill  \\
\end{array}} \right].
\label{eq:qlinear}
\end{equation}

From eq.~\eqref{eq:qlinear} it can be seen that the filter submatrices are shared among input dimensions, thus not only reducing the number of free parameters by $1/4$, resulting in a more lightweight model, but additionally endowing the neural network with the ability to grasp latent relations among channel dimensions. Nonetheless, QNNs are limited to $4$D inputs, therefore parameterized hypercomplex multiplication (PHM) \cite{zhang2021phm} and convolution (PHC) \cite{grassucci2021phnns} have been introduced to bridge this gap. The core idea of these methods lies in expressing the weight matrix as a sum of $n \in \mathbb{N}$ Kronecker products, thus we have

\begin{equation}
    \W = \sum_{i=0}^n \mathbf{A}_i \otimes \mathbf{F}_i,
\label{eq:phm}
\end{equation}

\noindent whereby matrices $\mathbf A_i$ encode the algebra rules, directly learned from the data, and $\mathbf F_i$ represent the filters. As a result of eq.~\eqref{eq:phm}, a parameterization of $\W$ is obtained, meaning that $n$ is a user-defined hyperparameter that decides in which domain the neural model operates (e.g., $n=4$ for the quaternion domain), thus extending the aforementioned properties of QNNs to general input domains $n$D. Specifically, PHM and PHC layers  employ $1/n$ free parameters with respect to real-valued counterparts and still possess the ability to model correlations present in the data, unlike real-valued networks.

\subsection{Multimodal Hypercomplex Fusion Network}

To address the challenges presented in Section~\ref{sec:background} we propose HyperFuseNet, a multimodal architecture that exploits hypercomplex algebra properties to effectively fuse the learned latent representations as can be seen in Fig.~\ref{fig:method}. The neural model comprises two main components, that is the encoder and a hypercomplex fusion module. Concretely, the encoder is composed of four different branches in the real domain, one for each modality, and has the objective of learning modality-specific latent representations directly from the raw signals, thus with a modality-level focus. Thereafter, these features in the latent space are merged together and processed by the proposed hypercomplex fusion module. The latter is composed of PHM layers with the hyperparameter $n$ set to $4$, as there are four feature vectors in input to the module corresponding to the four modalities, and has the role of learning a fused representation, thus performing a patient-level analysis. More in detail, by defining multiplications in the hypercomplex domain, the proposed fusion module possesses the capability of grasping cross-modal interactions between the learned latent features of the EEG, ECG, GSR, and eye data signals, which are highly correlated. Thus, the hypercomplex fusion module captures both global and local relations between feature dimensions, unlike real-valued networks, accordingly learning a more powerful representation by truly exploiting the correlations present in the different physiological signals.

\section{Experimental results}
\label{sec:exp}

% \begin{itemize}
%     \item Datasets description
%     \item Experimental setup
%     \item Description of validation metrics
%     \item First set of experiments
%     % \item Second set of experiments
%     % \item Ablation studies
% \end{itemize}

\subsection{Dataset}

To validate the proposed approach we adopt a publicly available dataset, that is MAHNOB-HCI \cite{soleymani2011mahnob}. It is a multimodal dataset for affect recognition which includes synchronized recordings of face video, audio signal, eye gaze data, and peripheral/central nervous system physiological signals of $27$ participants while watching emotional video clips. The eye gaze data comprises pupil dimensions, gaze coordinates, and eye distances, while for the physiological signals, we focus on EEG, ECG, and GSR, as these are highly related to emotional changes \cite{soleymani2011mahnob}. The database provides labels related to arousal, i.e., calm, medium aroused, and excited, and valence, i.e., unpleasant, neutral valence, and pleasant.

\subsection{Preprocessing and data augmentation}

% \begin{itemize}
%     \item sampling rates and downsampling
%     \item filters
%     \item baseline removal
%     \item keep eyes closed
%     \item eeg channel selection
%     \item sample extraction (10s windows) and data split
%     \item scaling + noise addition
% \end{itemize}

% \begin{table*}[t]
% \centering
% \caption{Results on MAHNOB-HCI of the proposed method compared against a state-of-the-art model with and without data augmentation.}
% \label{tab:results}
% \begin{tabular}{llccc}
% \toprule
% \multicolumn{1}{l}{Model} & \multicolumn{1}{c}{Params} & \multicolumn{1}{c}{Augm.} & \multicolumn{1}{c}{F1-score} & \multicolumn{1}{c}{Accuracy} \\  \midrule 

% Dolmans \cite{dolmans2021workload} & M & \multirow{2}{*}{\ding{55}} & 0. $\pm$ 0. & 0. $\pm$ 0. \\
% HypeFuse (ours) & M & & \textbf{0. $\pm$ 0.} & \textbf{0. $\pm$ 0.} \\ 
% \midrule
% Dolmans \cite{dolmans2021workload} & M & \multirow{2}{*}{\ding{51}} & 0. $\pm$ 0. & 0. $\pm$ 0. \\
% HypeFuse (ours) & M & & \textbf{0. $\pm$ 0.} & \textbf{0. $\pm$ 0.} \\ 
% \bottomrule   
% \end{tabular}
% \end{table*}

\begin{table*}[t]
\centering
\caption{Results on MAHNOB-HCI of the proposed method compared against a state-of-the-art model with and without data augmentation.}
\label{tab:results}
\begin{tabular}{@{}lccclcc@{}}
\toprule
\multirow{2}{*}{Model}  & \multirow{2}{*}{Augm.} & \multicolumn{2}{c}{Arousal} &  & \multicolumn{2}{c}{Valence} \\ \cmidrule(lr){3-4} \cmidrule(l){6-7} 
                                                    &                        & F1-score     & Accuracy     &  & F1-score     & Accuracy     \\ \midrule
Dolmans \cite{dolmans2021workload} & \multirow{2}{*}{\ding{55}}    & 36.60 $\pm$ 1.61  & \textbf{41.23} $\pm$ 2.03  &  & 37.44 $\pm$ 3.22  & 41.89 $\pm$ 3.34  \\
HyperFuseNet (ours)                                     &              & \textbf{38.83} $\pm$ 1.66  & 40.02 $\pm$ 1.98  &  & \textbf{41.43} $\pm$ 1.62  & \textbf{43.42} $\pm$ 2.57  \\ \midrule
Dolmans \cite{dolmans2021workload} & \multirow{2}{*}{\ding{51}}     & 38.86 $\pm$ 1.11  & 40.90 $\pm$ 0.62  &  & 38.33 $\pm$ 1.24  & 40.24 $\pm$ 1.04  \\
HyperFuseNet (ours)                                     &              & \textbf{39.65} $\pm$ 1.75  & \textbf{41.56} $\pm$ 1.33  &  & \textbf{43.60} $\pm$ 2.22  & \textbf{44.30} $\pm$ 2.01  \\ \bottomrule
\end{tabular}
\end{table*}

Firstly, we downsample EEG, ECG, and GSR signals from $256$Hz to $128$Hz, while we keep eye data at $60$Hz. Then, we filter EEG and ECG signals with a band-pass filter at $1$-$45$Hz and $0.5$-$45$Hz, respectively \cite{liu2017realtime, zeng2020elderly}, while a low-pass filter at $60$Hz is applied to GSR signals \cite{miranda2018amigos}, and for all of them an additional notch filter at $50$Hz \cite{maeng2020deep}, with all EEG signals being firstly referenced to average. Additionally, we perform a baseline correction on the GSR signal with respect to the mean value within the $200$ms preceding each trial to eliminate the initial offset of the signal. %as done automatically to the others by the high-pass filters
Finally, as for EEG data, we select $10$ channels out of the original $32$, i.e., F$3$, F$4$, F$7$, F$8$, FC$5$, FC$6$, T$7$, T$8$, P$7$, and P$8$, as these are the most related to emotion \cite{topic2022reduced, msonda2021channel}.
Instead, regarding eye data, we take the average between the signals related to the two eyes and we keep $-1$ values as they correspond to blinks or rapid movements which are relevant to the task at hand.
%The eye gaze data are kept at $60$Hz, but signals of the same type sampled from each eye are fused together by averaging. Whenever the data from both eyes are missing or incorrect (i.e., value outside possible limits), due for instance to blinks or rapid movements of the subject, a $-1$ replaces the measured value. We decided to keep these values in the signals, since blinks, voluntary eyes closure, and movements also are measures of interest for our task. Unfortunately, baseline correction is not applied because of the partial lack of baseline data in the dataset itself.
%non so se si capisce la parte in cui detto segnali dello stesso tipo fusi insieme facendo la media dei due occhi 

We extract samples by dividing the last $30$s of each trial into three segments of $10$s, as measurements toward the end of the clips reflect the emotion of the subjects' rating \cite{zeng2020elderly}. %Whenever a sample contains too many $-1$ values in the eye data (i.e., more than $40\%$ of the total segment length), it is discarded. 
Finally, we split the dataset in a stratified fashion by taking $20\%$ of the data for testing. Training samples are then augmented by applying scaling and noise addition. Firstly, two scaling factors are uniformly sampled over two intervals, i.e., $[0.7, 0.8]$ and $[1.2, 1.3]$, and applied to the original sample to generate two augmented versions. Then, a Gaussian noise signal with zero mean is added to each sample, with its standard deviation being computed modality-wise such that the augmented signal has a signal-to-noise ratio (SNR) of 5dB. A total of $30$ augmented signals are generated for each original sample. %In both cases, the $-1$ values in the eye data are left untouched.
%non spiegato che per eye va fatto downsampling del noise, entriamo troppo nel tecnico forse

\subsection{Architecture and training recipe}
\label{subsec:training}

The proposed architecture comprises four branches that compose the encoder and a hypercomplex fusion module. The branches consist of three fully-connected layers, except for the GSR branch which has two, with $128$ units for eye data and GSR, $512$ for ECG, and $1024$ for EEG, interleaved with batch normalization and ReLU activation function, inspired by \cite{dolmans2021workload}. Then, the learned latent representations are merged together and processed by the proposed fusion module which comprises four PHM layers with $n=4$, with the same interleaved layers and the number of units halved at each layer, a dropout layer, and the final output layer. The model is trained using the Adam optimizer, with a categorical cross-entropy loss and a one-cycle policy. The best hyperparameters are found by doing a bayesian search, sampling the learning rate from $[0.001, 0.008]$. The number of epochs is set to $100$ with early stopping with patience at $20$.

\subsection{Results}

We report in Tab.~\ref{tab:results} the results of the conducted experimental analysis, showing the mean over $3$ runs of the F1-score and accuracy, which indeed is not always representative due to imbalance of classes. In detail, we compare the proposed architecture against a state-of-the-art multimodal network that also operates with raw signals and is originally designed for mental workload classification. We train it using the same approach we employed for our network on the same database. Firstly, we can observe that the employed data augmentation is effective and improves the performance of both networks. Secondly, and most importantly, the proposed hypercomplex architecture outperforms the method employed as comparison in both augmentation scenarios, thus demonstrating the efficacy of the PHM layers in the fusion step which yield better emotion recognition accuracy as a result of the grasped cross-modal correlations thanks to hypercomplex algebra rules.

\section{Conclusion}

In this paper, we proposed a multimodal architecture with a novel hypercomplex fusion module for emotion recognition from EEG and peripheral physiological signals, in which a modality-specific representation is firstly learned in the real domain and consequently processed together by the fusion module in the hypercomplex domain. The latter was found to be effective to perform a more proper fusion step than classical real-valued fully-connected layers, in fact, by employing hypercomplex multiplications the module is capable of capturing relations among the learned latent features and as a result learn a more discriminant representation. In future efforts, we aim at additionally exploiting intra-modality correlations with parameterized hypercomplex convolutions, thus bringing the advantages of the fusion step also at the encoder level.

\bibliographystyle{IEEEtran}
\ninept
\bibliography{HEEG.bib}

\end{document}